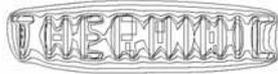
Budapest, Hungary, 17-19 September 2007# New Reliability Assessment Method for Solder Joints in BGA Package by Considering the Interaction between Design Factors

Satoshi KONDO, Qiang YU, Tadahiro SHIBUTANI, Masaki SHIRATORI

Department of Mechanical Engineering and Materials Science, Yokohama National University
79-5, Tokiwadai, Hodogaya-ku, Yokohama 240-8501, Japan
Phone: +81-45-339-3862 Fax: +81-45-331-6593, E-mail: qiang@swan.me.ynu.ac.jp*Abstract*- As the integration and the miniaturization of electronics devices, design space become narrower and interactions between design factors affect their reliability. This paper presents a methodology of quantifying the interaction of each design factor in electronics devices. Thermal fatigue reliability of BGA assembly was assessed with the consideration of the interaction between design factors. Sensitivity analysis shows the influence of each design factor to inelastic strain range of a solder joint characterizing the thermal fatigue life if no interaction occurs. However, there is the interaction in BGA assembly since inelastic strain range depends on not only a mismatch in CTE but also a warpage of components. Clustering can help engineers to clarify the relation between design factors. The variation in the influence was taken to quantify the interaction of each design factor. Based on the interaction, simple evaluating approach of inelastic strain range for the BGA assembly was also developed. BGA package was simplified into a homogeneous component and equivalent CTE was calculated from the warpage of BGA and PCB. The estimated equation was derived by using the response surface method as a function of design factors. Based upon these analytical results, design engineers can rate each factor's effect on reliability and assess the reliability of their basic design plan at the concept design stage.## I. Introduction

The electronic product manufacturing industry encounters strong competition from short cycle time manufacturing and shrinkage of cost in recent years. Design process, in particular at an early stage, is the key to reduce waste and cycle time in manufacturing. The miniaturization and the high integration of electronic devices were also progressed by the advance in technology. Furthermore, the reliability of fatigue life has been prioritized as an important concern, because the thermal expansion difference between a package and printed circuit board causes thermal fatigue. In addition, design factors have the interaction because of the complex structure. The reliability engineers have to remedy this reliability problem in early design stage, but it is very difficult and needs much cost because of the interaction between design factors.

In the study, new evaluating method of the thermal fatigue reliability for solder joints in electronic devices was developed. The method of understanding the relation between design factors and the thermal fatigue life has been established, and an application of the BGA (Ball Grid Array) package was examined. As a result, the interaction between design factors was clarified. Based on the interaction, a simple evaluation technique of the thermal fatigue reliability in early design stage was developed.

In order to improve the reliability, design engineer needs understand the relation between each design factors and characteristic value like inelastic equivalent strain range in the solder joint for thermal fatigue reliability. To clarify the main effects and the interaction of each design factors on reliability, sensitivity analysis and cluster analysis were executed. As a result, the influence of the interaction between each design factor has been clarified, and this result indicates that the warpage of package is very important issue for assessing the reliability and the key design characteristic linking with the non-linear equivalent strain range.

Based upon the results of sensitivity and interaction studies, it is shown that the non-linear equivalent strain range of solder joints can be expressed with high accuracy by very simple equation. And then, the reliability of BGA solder joints can be directly assessed by this simple equation. This approach of reliability assessment is called as simple assessment approach in this study.

This result would lead to the understanding of the relation between each design factor and the thermal fatigue life. And it is possible to evaluate the thermal fatigue reliability simply at the early stage of the design development before generating detailed analytical model or the fatigue test of the real component. Therefore, this result will help for an adequate design.

### NOMENCLATURE

FEM  Finite Element Method
BGA  Ball Grid Array
PCB  Printed Circuit Board
CTE  Coefficient of Thermal Expansion
DOE  Design of Experiment

©EDA Publishing/THERMINIC 2007       -page-       ISBN: 978-2-35500-002-7

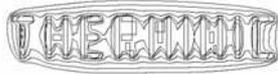

## II. SENSITIVITY ANALYSIS

Thermal fatigue of solder joints is a primary failure mechanism of a BGA assembly. Several methods can be used to evaluate the thermal fatigue life of solder joints [1-2]. It is known that the total equivalent inelastic strain range per cycle of thermal load can be used to evaluate the fatigue life based on Manson-Coffin law [3-6]. Then, the initial fatigue crack occurring in solder joints is used for the thermal fatigue life [7-8].

Sensitivity analysis was carried out to clarify the relation between a set of design factors and the total equivalent inelastic strain range. At first, sizes, mechanical properties and thermal properties of components were taken as the design factors from a BGA assembly model shown in Fig.1. Each design factor has three levels listed in Table 1. A set of 27 models for the case study was designed by using DOE theory. In this case study, the total equivalent inelastic strain range of the solder joint is the characteristic value calculated by using FEM analysis. The other detail conditions of FEM analysis are stated as follows:

1) The temperature range is from -40ºC to +125ºC for the thermal load. The temperature changing time is 0.05 hour (3 minutes), and the dwelling time is 0.25 hour (15 minutes) as shown in Fig. 2.

2) Based upon the symmetry of the package structure, a quarter of the BGA model was used in this analysis, and symmetrical boundary conditions are subjected on the boundaries shown in Fig.3.

3) The total equivalent inelastic strain range can be calculated by the average values of each node around 50μm area at the corner of the solder bump.

Using response surface method, the characteristic value can be expressed with the basis of estimated equation. And the influence of each design factor on the total equivalent inelastic strain range was calculated. The influence ratios of design factors are shown in Table.2. The influence figure of each design factor was also shown in Fig. 4. Based upon these results, the design engineers can roughly identify how much influence each design factor has on the characteristic value.

By this analytical result, a very rough relation between a very narrow design change of each design factor and the thermal fatigue life become clearer, and design engineers understand the effect of their minor design change on the reliability. However, if a big design change is needed on a design change causes a different balance of the component, the results of DOE study will lose the function for estimating the design change impact on the reliability, because the interaction relation between design factors could not be included in the influence analysis on the characteristic. Furthermore, since only very limited capacity in DOE analysis can be used to search and study the interaction relation between design factors, authors proposed a new approach by clustering analysis to cover this kind of design support issue.

## III. CLUSTER ANALYSIS

To clarify the interaction of each design factor, the cluster analysis was used in this study. This approach consists of three basic processes, the parametric-study function, grouping function for the similar design cases, and analytical function for the relation between design input and characteristics. In order to group the models by each feature based on the characteristic of the analyses, hierarchical cluster analysis [9-12] was performed. In order to show a clear trend of the relation between design factors and characteristics, the average value of each design factor was calculated and is shown in the order with performance level of the design characteristics. Furthermore, the interaction [13-14] between design factors are taken in every cluster, and the average of the interaction coefficients for all clusters is calculated, so that relation between design factors could be simply caught. Cluster Analysis is the method of calculating the Euclid distance between parameters, and gathering close models in proper group, and expressing group's relation by a hierarchical structure. Hierarchical cluster analysis was performed based on the values of non-linear strain range. Every cluster was judged that the characteristic value has been gathered by every feature based on the Euclid Distance, where the design characteristic was expressed by the non-linear strains in this study.

The values of the total equivalent inelastic strain range obtained from FEM analysis based on the orthogonal table were plotted in Fig. 5. In this figure, the design numbers are shown in order of the strain range value, and as a result of the design cases can be clustered into 4 groups by the value of the strain range. Here the groups are called as the clusters. In order to investigate the relation between each design factor, design factors in each cluster were averaged and the average values of design factors in each cluster are shown in Fig. 6, where the value of each design factor was regularized by design range. 1.0 means the maximum value in this study for every design factor, and then 0.0 means the minimum one used in the case study. As shown in Fig.6, the average values of design factors plotted as No.1cluster shows the basic design pattern or trend which will achieve the designs with the lowest strain range in the solder joints. On the other hand, the pattern plotted as No.4 will give cause the biggest strain range and lowest fatigue life. The arrows show the trends on total equivalent inelastic strain range, and its length indicates the intensity of influence to the characteristic value.

From this figure, as the thickness and CTE of encapsulant decrease, and the thickness and CTE of substrate increase, the total equivalent inelastic strain range tends to decrease. In the case of this condition, it is easily estimated that the whole package would curve upward in high temperature. And the mismatch in shear direction between the component and PCB tends to decrease when the package warps upwards as shown in Fig. 7 [15]. Therefore, it is clarified that the warpage of package has a large influence to the reliability. The influence of the warpage must be considered in package design stage, and the consideration of the warpage is essentially same as the consideration of interaction between each design factor.

This cluster analysis clarifies the whole interaction of each design factor. Furthermore, in order to achieve a more exact understanding, it is important to investigate more detail interaction of design factors.

IV. ANALYSIS OF INTERACTION BETWEEN DESIGN FACTORS

A basic idea is that the interaction between a certain design factor A and the others can be checked by investigating the influence on design results from all design factors when the value of factor A is fixed. If the impact on the design results from a design factor's change is dependent on the value of A, it means there is an interaction relation between the design factor and A. This process can be carried out very easily by using the existing data for the case studying shown in Fig.5. At first, the cases shown in Fig.5 were grouped into two sets by the value of the substrate thickness. One group includes the cases with the maximum substrate thickness, and the other consists of the cases with the thinnest substrate. And then, clustering analysis was carried out by the value of non-linear strain range for two groups respectively, and the results are shown in Fig. 8. Here two clusters were formed for each group, 1 and 2 show the clusters with the lower non-linear strain range when the substrate thickness was fixed at minimum and maximum values respectively, and 1' and 2' show the clusters with the higher non-linear strain range.

Based upon results of clustering analysis, the average value of each design factor was calculated within each cluster, and the results are shown in Fig. 9. This figure shows how the substrate thickness value affects the other design factor's trends which link the correlations between the design factors and the strain ranges.

For example, when the thickness of substrate is low, the thickness of chip should be kept lower in order to decrease the total equivalent inelastic strain range. But when the thickness of substrate is high, the chip thickness should be higher. It turned out that when one design factor (in this case, the thickness of substrate) was changed, the influence trend of another design factor (thickness of chip) on the characteristic value was reversed. It means there is a strong interaction relation between the two design factors. In the similar case, when the thickness of substrate is low, the CTE of substrate and the CTE of encapsulant don't affect the characteristic value. But when the thickness of substrate is high, these factors affect to the characteristic value remarkably. Therefore, it became clearer that the thickness of chip, the CTE of substrate and the CTE of encapsulant have strong interaction with the thickness of substrate. And the coefficient of interaction can be calculated from the difference between regularized values (arrows in the figure).

By applying this clustering technique for all design factors, the coefficient of interaction between all design factors was clarified. All coefficient of interaction is shown in Table.3. From this result, it is clarified that there is the interaction between the PCB and the BGA package.

V. SIMPLE RELIABILITY ASSESSMENT APPROACH CONSIDERING WARPAGE OF PACKAGE

From the previous sections, some design factors of BGA package have strong interaction between each other, which causes the warpage of package. Since the warpage affects the mismatch in deformation between the package and the circuit board, inelastic strain range in solder joints should be assessed considering the interaction of design factors. Then, an equivalent CTE of BGA package can be easily calculated by considering the effect of the package's warpage as shown in Fig. 10. Thermal mismatch generated in solder joints can be simplified into a mismatch in CTE. The difference of OA-OB with $\Delta T$ is equivalent to expansion due to equivalent CTE.

By using this method, the reliability of various packages with complex structure would be assessed. And this method becomes important because of the advance of recent chip stacking technology.

However, from the cluster analysis, the factors of the PCB also show the complex interaction relation. It means that the warpage of package gave great effect not only on the behavior of the package itself but also on the mounded structure or the equivalent thermal mismatch between the package and the PCB. To solve this problem, not only the equivalent CTE of the whole package, but also the equivalent thermal mismatch between the package and PCB should be assessed by considering the influence of the warpage.

The behavior of the warpage depends on the relative rigidity of components. When the bending rigidity of PCB is larger than that of the package, the warpage of package might be suppressed to the PCB during mounting process. On the other hand, when the PCB is thin, the warpage of PCB also occurs due to the BGA warpage. Therefore, it is necessary to correct this CTE according to the ratio of the bending rigidity of the package and PCB.

Here the curvature radius of the package ($1/R_{pre}$) was calculated from the result shown in Fig. 10. After mounting, the PCB and package would be suppressed to each other and the package and PCB would share the same curvature radius as shown in Fig. 11. The curvature, $1/R_{new}$ after the mounting as follows:

$$\frac{1}{R_{new}} = \frac{1}{R_{pre}} \frac{EI_z^{BGA}}{EI_z^{BGA} + EI_z^{PCB}} \qquad (1)$$

Here, $E$ and $I_z$ are Young's modulus and the second moment of area. Superscript indicates each component. From the curvature radius after mounting, new equivalent CTE of the package and PCB were calculated from the same method as Fig. 10. As a result, the equivalent CTE includes the more accurate influence of the warpage.

Using these equivalent CTE, simplified three-layer model in Fig. 12 (homogenous package, solder joints and PCB) can be used for evaluating the inelastic strain range. Equivalent CTE and Young's modulus of whole package were defined as material property of the homogeneous package. Then FEM analysis was executed, and the result of simple reliability assessment was compared from the result of detail BGA model analysis as shown in Table.4. If the response surface is constructed without the influence of PCB, the error of estimated value is over 20 %. It was obvious that the proposed simple model with PCB influence has higher accuracy. The simple reliability assessment becomes more accurate by considering the warpage of package. This simplified model is useful for design process in the early stage and the interaction between design factors is also useful for design engineers.

## VI. CONCLUSIONS

To investigate the interaction in BGA assembly, the sensitivity of each design factor and the interaction between design factors were analyzed. Based on the warpage from the interaction in the BGA package, the simple approach to evaluate the thermal fatigue life was developed. Obtained results can be summarized as follows:

1) The sensitivity analysis shows the influence of each design factor on thermal fatigue life without the consideration of the interaction. However, because of the interaction between design factors, this method will lose the function.
2) Clustering technique can provide the whole interaction of each design factor, and it is clarified that the warpage of package has large influence.
3) Moreover, by clustering for the observed design factor, detail interaction of each design factor can be extracted. And it is clarified that there is the interaction between BGA package and PCB.
4) By considering the warpage of package and the interaction between the package and PCB, it became possible to assess the reliability of BGA package easily in high accuracy.

By this study, design engineers can rate each factor's effect on the reliability and assess the reliability of their basic design plan at the concept design stage.

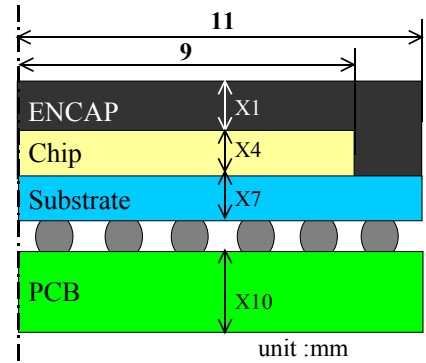

Fig. 1 BGA Package model

Table 1. BGA Package design factors and design range

|     | Design factors | Min. | Ave. | Max. |
| --- | --- | --- | --- | --- |
| X1 | Thickness of Chip (μm) | 300 | 400 | 500 |
| X2 | Thickness of Sub' (μm) | 300 | 400 | 500 |
| X3 | Thickness of PCB (μm) | 800 | 1000 | 1200 |
| X4 | Thickness of Encap (μm) | 1000 | 1200 | 1400 |
| X5 | Young's modulus of Sub' (GPa) | 15 | 19 | 23 |
| X6 | CTE of Sub' ($10^{-6}$/°C) | 12 | 15 | 18 |
| X7 | Young's modulus of PCB (GPa) | 15 | 19 | 23 |
| X8 | CTE of PCB ($10^{-6}$/°C) | 13 | 16 | 19 |
| X9 | Young's modulus of Encap (GPa) | 13 | 16 | 19 |
| X10 | CTE of Encap ($10^{-6}$/°C) | 12 | 15 | 18 |

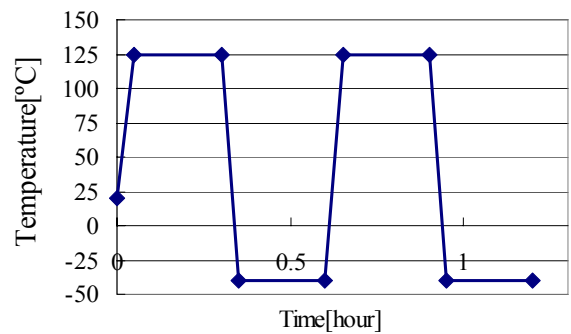

Fig. 2. Thermal load of analysis

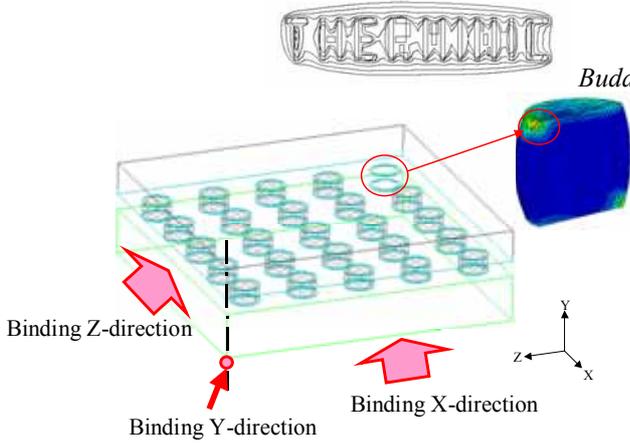

Fig. 3. Boundary conditions

Table 2. Influence levels of design factors BGA package

| Factor | Effective ratio |
|---|---|
| Height of Chip | 0.39% |
| Height of Sub' | 9.43% |
| Height of PCB | 0.03% |
| **Height of Encap** | **38.90%** |
| Young's modulus of Sub' | 9.37% |
| **CTE of Sub'** | **19.81%** |
| Young's modulus of PCB | 0.00% |
| CTE of PCB | 7.94% |
| Young's modulus of Encap | 2.03% |
| CTE of Encap | 9.63% |
| Error | 2.44% |
| Total | 100.00% |

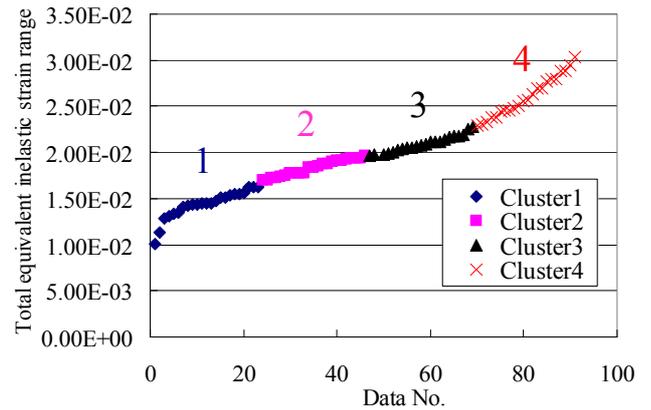

Fig. 5. Data of total equivalent inelastic strain range

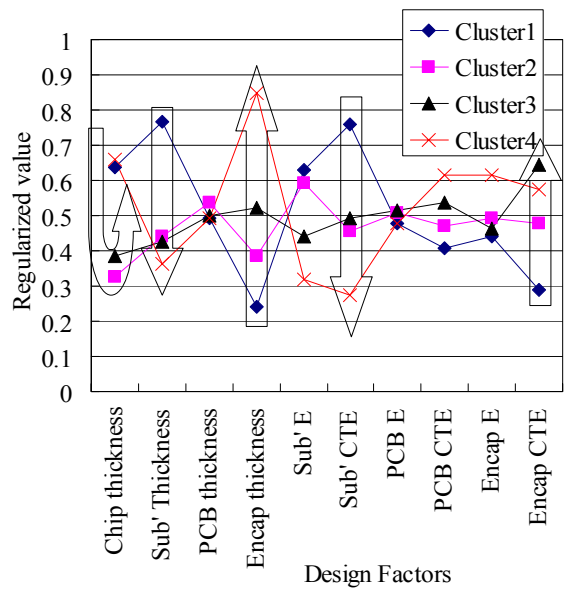

Fig.6. Clustering of BGA package design factor

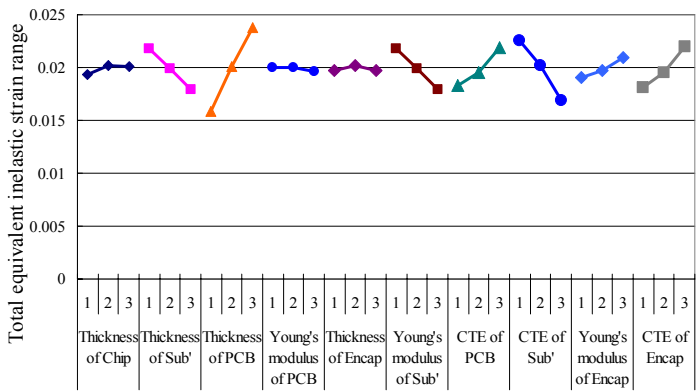

Fig. 4. Influence figure of design factors

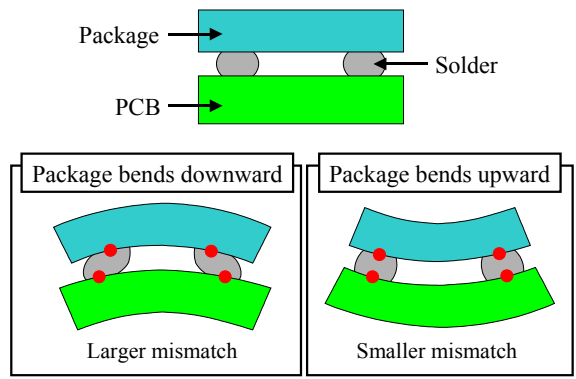

Fig.7. Structure of influence in the solder by curvature

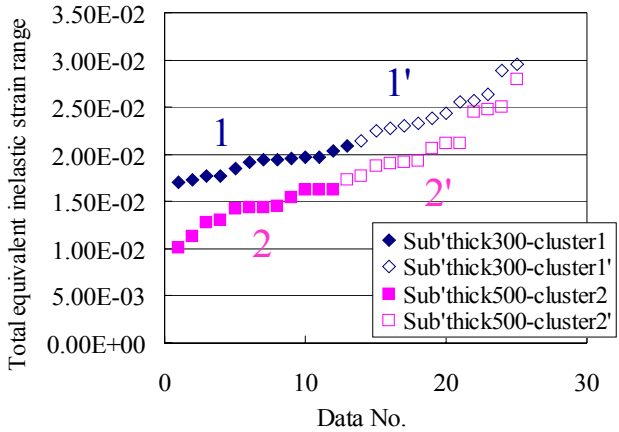

Fig. 8. Data of total equivalent inelastic strain range in cases of substrate thickness is maximum and minimum

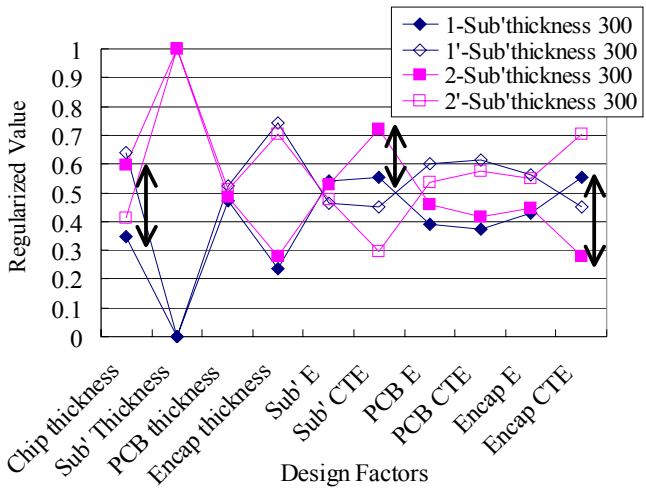

Fig. 9. Clustering result when substrate thickness changed

Table. 3. Matrix of interaction coefficient between all design factors

|  | Chip Thickne | Sub' Thickn | PCB thickne | Encap thickne | Sub' E | Sub' CTE | PCB E | PCB CTE | Encap E | Encap CTE |
|---|---|---|---|---|---|---|---|---|---|---|
| Chip thickness | - | 0.19 | 0.01 | -0.15 | 0.19 | 0.10 | 0.07 | 0.03 | -0.15 | 0.21 |
| Sub' Thickness | 0.25 | - | 0.01 | 0.04 | -0.01 | 0.17 | 0.07 | 0.04 | 0.01 | -0.28 |
| PCB thickness | -0.11 | 0.03 | - | 0.11 | 0.15 | 0.00 | 0.19 | 0.04 | 0.28 | -0.03 |
| Encap thickness | -0.37 | 0.01 | 0.06 | - | 0.06 | -0.01 | -0.13 | 0.00 | 0.01 | 0.00 |
| Sub' E | 0.43 | 0.07 | 0.10 | 0.04 | - | 0.06 | -0.14 | 0.04 | 0.04 | 0.14 |
| Sub' CTE | 0.15 | 0.25 | -0.11 | 0.14 | 0.01 | - | -0.01 | 0.22 | 0.04 | -0.32 |
| PCB E | 0.01 | 0.25 | 0.21 | 0.08 | -0.01 | -0.03 | - | 0.18 | 0.14 | 0.11 |
| PCB CTE | -0.04 | 0.04 | 0.13 | 0.13 | 0.10 | 0.31 | 0.19 | - | 0.13 | 0.33 |
| Encap E | -0.14 | -0.08 | 0.29 | 0.13 | 0.14 | 0.08 | 0.15 | -0.07 | - | -0.11 |
| Encap CTE | -0.13 | -0.25 | -0.03 | -0.08 | 0.28 | -0.32 | -0.04 | 0.24 | -0.25 | - |

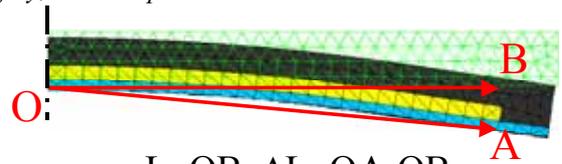

$$L=OB, \Delta L=OA-OB$$

$$CTE = \frac{\Delta L}{L \cdot \Delta T}$$

Fig. 10 The calculation method of whole package CTE

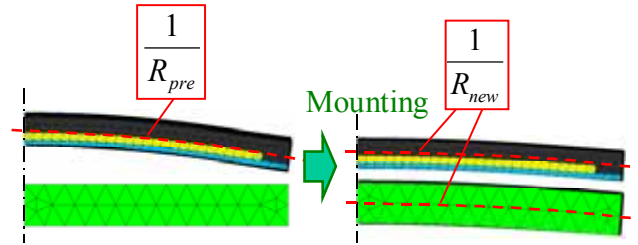

Fig. 11. The correction of curvature radius

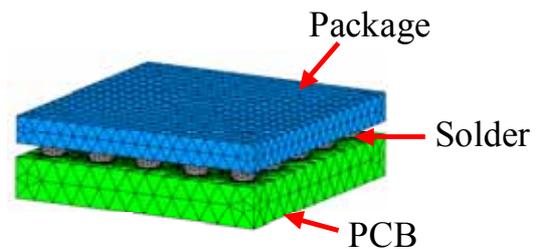

Fig. 12. Three-layer model

Table. 4. Accuracy of simple reliability assessment approach

| Design Condition | |
|---|---|
| Encap CTE [$10^{-6}/°C$] | 18 |
| Encap Young's Modulus [MPa] | 17000 |
| Encap thickness [mm] | 0.3 |
| Chip CTE [$10^{-6}/°C$] | 3 |
| Chip Young's Modulus [MPa] | 160000 |
| Chip thickness [mm] | 0.1 |
| Sub' CTE [$10^{-6}/°C$] | 14 |
| Sub' Young's Modulus [MPa] | 20000 |
| Sub' thickness [mm] | 0.2 |
| PCB thickness [mm] | 1.1 |
| PCB Young's Modulus [MPa] | 19000 |
| PCB CTE [$10^{-6}/°C$] | 19 |

| | $\Delta\varepsilon_{in}$ | Error margin |
|---|---|---|
| Detail model | 5.81E-03 | - |
| Simple model (no pcb influence) | 7.16E-03 | 23.4% |
| Simple model (pcb influence considered) | 6.14E-03 | 5.8% |